\documentclass[letterpaper, 10pt, conference]{ieeeconf}

\usepackage[T1]{fontenc}
\usepackage[utf8]{inputenc}

\usepackage{amsmath}
\usepackage{amsfonts}
\usepackage{gensymb}
\usepackage[per-mode=symbol]{siunitx}
\usepackage{cite}
\usepackage{booktabs}
\usepackage[capitalise]{cleveref}
\crefname{subsection}{subsection}{subsections}
\crefname{problem}{Problem}{Problems}
\crefname{equation}{}{}
\newtheorem{problem}{Problem}
\newtheorem{remark}{Remark}
\usepackage{cases}

\usepackage{graphicx}
\usepackage{cuted}
\usepackage{xcolor}
\definecolor{vgRed}{RGB}{193, 48, 24}
\definecolor{vgOrange}{RGB}{243, 111, 19}
\definecolor{vgYellow}{RGB}{235, 203, 56}
\definecolor{vgGreen}{RGB}{162, 185, 105}
\definecolor{vgLightBlue}{RGB}{13, 149, 188}
\definecolor{vgDarkBlue}{RGB}{6, 56, 81}

\usepackage[acronym,shortcuts]{glossaries}
\newacronym{bem}{BEM}{blade element momentum theory}
\newacronym{mpc}{MPC}{model predictive control}
\newacronym{mibp}{MIBP}{mixed integer bilinear program}
\newacronym{mimo}{MIMO}{multi-input-multi-output}
\newacronym[longplural=damage equivalent loads]{del}{DEL}{damage equivalent load}
\newacronym{rms}{RMS}{root mean square}
\newacronym{lidar}{LIDAR}{light detection and ranging}

\usepackage{todonotes}
\renewcommand{\todo}[2][]{\tikzexternaldisable\@todo[#1]{#2}\tikzexternalenable}

\usepackage{tikz}
\usetikzlibrary{calc}
\usetikzlibrary{decorations.pathreplacing,decorations.markings}
\usepackage{pgfplots}
\pgfplotsset{compat=1.15}
\usepgfplotslibrary{external}
\usepgfplotslibrary{statistics}
\tikzset{external/system call={pdflatex \tikzexternalcheckshellescape -halt-on-error
    -interaction=batchmode -jobname "\image" "\texsource"}}
\pgfplotsset{every axis/.append style={semithick,tick style={major tick
            length=4pt,semithick,black}}}

\pgfplotsset{
    colormap={customColorMap}{
        rgb255=(6, 56, 81)
        rgb255=(13, 149, 188)
        rgb255=(162, 185, 105)
        rgb255=(235, 203, 56)
        rgb255=(243, 111, 19)
        rgb255=(193, 48, 24)
    },
}

\pgfplotsset{
    boxplot/whisker range = {1000},
    boxplot/box extend = 0.4,
}

\pgfkeys{/pgfplots/x axis shift down/.style={
        x axis line style={yshift=-#1},
        xtick style={yshift=-#1},
        tick align=outside,
        xticklabel shift={#1}}}

\pgfkeys{/pgfplots/y axis shift left/.style={
        y axis line style={xshift=-#1},
        ytick style={xshift=-#1},
        yticklabel shift={#1},
        scaled y ticks = false,
        tick align=outside,
        y tick label style={/pgf/number format/fixed,
        }
    }
}

\pgfplotsset{myPlot/.style={%
        width=8cm,
        height=4cm,
        line width = 0.6pt,
        separate axis lines,
        axis x line*=bottom,
        x axis shift down = 3pt,
        enlarge x limits=false,
        axis y line*=left,
        y axis shift left = 6pt,
        enlarge y limits={abs=.25pt},
        enlarge x limits={abs=.25pt},
    }
}

\pgfdeclareshape{lowpassnode}
{
    \inheritsavedanchors[from={rectangle}]
    \inheritbackgroundpath[from={rectangle}]
    \inheritanchorborder[from={rectangle}]
    \foreach \x in {center,north east,north west,north,south,south east,south west, west, east}
    {
        \inheritanchor[from={rectangle}]{\x}
    }
        \foregroundpath
        {
            \pgfpointdiff{\northeast}{\southwest}
            \pgf@xa=\pgf@x \pgf@ya=\pgf@y
            \northeast
            \pgfpathmoveto{\pgfpoint{0.4\pgf@xa}{-0.2\pgf@ya}}
            \pgfpathlineto{\pgfpoint{0}{-0.2\pgf@ya}}
            \pgfpathlineto{\pgfpoint{-0.4\pgf@xa}{0.2\pgf@ya}}
            \pgfusepath{stroke}
        }
}

\newcommand{\gettikzxy}[3]{%
  \tikz@scan@one@point\pgfutil@firstofone#1\relax
  \edef#2{\the\pgf@x}%
  \edef#3{\the\pgf@y}%
}

\newcommand{\Rp}{\mathbb{R}^+}
\newcommand{\R}{\mathbb{R}}

\newcommand{\llambda}[1]{R\frac{\omega_r(#1)}{V(#1)}}

\IEEEoverridecommandlockouts
\title{\LARGE \bf
   Model Predictive Control of Wind Turbines with Piecewise-Affine Power Coefficient Approximation
}

\author{Arnold Sterle, Aaron Grapentin, Christian A. Hans and Jörg Raisch
\thanks{A. Sterle, A. Grapentin, J. Raisch and C. A. Hans are with Technische Universität Berlin, Control Systems Group, Germany \texttt{\{grapentin, sterle, raisch, hans\}@control.tu-berlin.de} J. Raisch is also with Science of Intelligence, Research Cluster of Excellence.}%
\thanks{This work was partially supported by the German Federal Ministry for Economic Affairs and Climate Action (BMWK), project No. 03EE2036C.}%
}

\begin{document}
\maketitle

\begin{abstract}

In this paper, an offset-free bilinear \acl{mpc} approach for wind turbines is presented.
State-of-the-art controllers employ different control loops for pitch angle and generator torque which switch depending on wind conditions.
In contrast, the presented controller is based on one unified control law that works for all wind conditions.
The inherent nonlinearity of wind turbines is addressed through a piecewise-affine approximation, which is modelled in a mixed-integer fashion.
The presented controller is compared to a state-of-the-art baseline controller in a numerical case study using OpenFAST.
Simulation results show that the presented controller ensures accurate reference power tracking.
Additionally, damage equivalent loads are reduced for higher wind speeds.

\end{abstract}


\section{INTRODUCTION}

The global goal of mitigating climate change remains a driving force for renewable energy sources.
To achieve lower greenhouse gas emissions, renewable energy resources have to be widely deployed \cite{edenhofer2011ipcc}.
The globally installed wind power capacity has been increased to $\SI{837}{\giga\watt}$ in 2021 with $\SI{72}{\percent}$ being installed in China, the US, Germany, India and Spain \cite{global2021global}.
On top, the cost of renewable energy has significantly declined over the last decade \cite{IRENA2021}.
This trend is expected to continue in the upcoming years.
The long-term operation and maintenance costs consist mainly of replacement costs \cite{vachon2002long}.
Therefore, reducing wear becomes a key factor for the future development of wind power.

Two modes of wind turbine operation can be considered depending on wind conditions. At lower wind speeds, generated power shall be maximized. At higher wind speeds, generated power shall be curtailed such that a given reference power $P_e^d$ is tracked, which is referred to as power tracking.
Traditionally, control laws with individual loops for pitch and torque have been employed.
These are switched depending on wind conditions (see, e.g., \cite{Aho2012,Kim2018,Apata2020}).
A multivariable control approach such as \ac{mpc} may improve control performance by considering pitch and torque interactions.
However, formulating models and objectives that hold for large wind speed ranges remains a challenging objective which has previously been dealt with by switching cost functions \cite{Schlipf2012,Mirzaei2013} or linear parameter varying models \cite{Morsi2015}.
Potentially, such approaches may suffer from reduced control performance at operations close to the point of switching.
Also, linear parameter varying models need to be carefully designed to properly reflect the system dynamics for the entire operation range.

In this paper, an offset-free \ac{mpc} scheme that is based on a mixed-integer bilinear program is presented.
The controller combines pitch and torque control into a unified control law that holds for all wind conditions such that no switching is required.
Additionally, it achieves improved power maximization capabilities while reducing loads at higher wind speeds.
Our \ac{mpc} scheme is thoroughly tested against a state-of-the-art controller in comprehensive numerical simulation case studies that employ OpenFAST \cite{openFAST}.
Our \ac{mpc} scheme proves to ensure accurate power tracking while reducing damage equivalent loads at higher wind speeds.

The paper is organized as follows.
In \cref{sec:model}, a state model of a wind turbine is derived.
Then, the \ac{mpc} scheme is formulated in \cref{sec:mpc}.
In \cref{sec:caseStudy}, simulation results of the \ac{mpc} approach and the baseline controller are presented and compared.
Finally, conclusions are drawn in \cref{sec:conclusion}.

\subsection{Preliminaries}\label{subsec:notation}

The sets of positive integers, integers and real numbers are denoted by $\mathbb{N},\mathbb{Z}$ and $\R$, respectively.
The set of positive real numbers is denoted by $\Rp$.
The set of nonnegative real numbers is denoted by $\Rp_0$.
Inequalities between vector quantities are evaluated element-wise.


\section{MODEL}
\label{sec:model}
Similar to previous work (see, e.g., \cite{Morsi2015,Merabet2011}), we use a wind turbine model that is composed of three subsystems:
aerodynamics, drive train and generator.
The aerodynamic system, which consists of the wind turbine's rotor hub and blades, converts wind into rotational motion.
The drive train's gearbox converts the rotor's slow speed into the generator's fast one.
Finally, the generator converts mechanical into electrical power.
In what follows, these parts will be discussed in detail.

\subsection{Aerodynamic subsystem}
\begin{figure}[t]
	\centering

\tikzset{external/export next=false}
\begin{tikzpicture}[font=\scriptsize]
  \begin{axis}[%
    xlabel = {$\lambda$},
    ylabel = {$u_\theta$ in $^\circ$},
    zlabel = {$C_p(\lambda, u_\theta)$},
    view = {20}{-30},
    axis lines = middle,
    xlabel style={anchor=south},
    ylabel style={anchor=west},
    zlabel style={anchor=south},
    xmax = 14.9,
    xmin = 1,
    ymax = 29,
    ymin = 0,
    zmax = 0.5,
    zmin = 0,
    restrict z to domain = 0:1,
    restrict y to domain = 0:26,
    restrict x to domain = 0:12,
    height = 5.6cm,
    width=8cm,
    ]

    \addplot3[surf, mesh/cols=20, z buffer = auto, opacity=0.5] table [x=lambda, y=theta, z=Cp, col sep=comma] {figures/csvFiles/powerCoefficientData.csv} ;

  \end{axis}
\end{tikzpicture}
	\caption{Power coefficient $C_p\left(\lambda,u_\theta\right)$ (data from \cite{RWT}).}
	\label{fig:powerCoefficientData}
\end{figure}
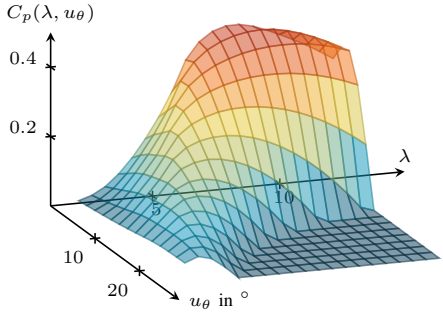

Wind passing through the rotor swept area ${\pi R^2}$, with radius ${R\in\Rp}$, exhibits a power of
\begin{equation}
P_{wind}(t) = m'V(t)^3,\label{eq:windPower}
\end{equation}
with ${m' = \frac{1}{2}\rho\pi R^2}$ where ${\rho\in\Rp}$ denotes the air density and ${V(t)\in\Rp}$ the rotor effective wind speed at time $t\in\R$ \cite[ch. 1]{Hansen_2015}.
For the case of $V(t) = 0$, the wind turbine is not operated.
Hence this case is disregarded here.
According to Betz's law \cite[ch. 1]{Hansen_2015}, the rotor can only convert a fraction of the wind power into rotational power ${P_r(t)\in\R}$.
This fraction is referred to as power coefficient $C_p\in\R$ which depends on various parameters, such as rotor blade shape, material and weight.
For a given turbine topology, it further depends on the wind speed's angle of attack on the rotor blades which changes with the blades' pitch angle $u_\theta(t)\in\R$ and the tip-speed-ratio $\lambda(t)\in\Rp$ which is given by
\begin{equation}
\lambda(t)=\llambda{t}\label{eq:tipSpeedRatio},
\end{equation}
where ${\omega_r(t)\in\Rp}$ denotes the rotor's angular velocity.
For the case of $\omega_r(t)=0$, the wind turbine is not operated.
Hence this case is disregarded here.
With $u_\theta(t)$ and $\lambda(t)$, the power coefficient can be defined as
\begin{equation}
	C_p\left(\lambda(t), u_\theta(t)\right) = \frac{P_r(t)}{P_{wind}(t)}.\label{eq:powerCoefficient}
\end{equation}
\Cref{fig:powerCoefficientData} depicts the power coefficient of a ${\SI{3.4}{\mega\watt}}$ wind turbine for different values of $\lambda$ and $u_\theta$.

Combining \eqref{eq:windPower} and \eqref{eq:powerCoefficient} yields
\begin{equation}
P_r(t) = m'V(t)^3C_p(\lambda(t),u_\theta(t))\label{eq:rotorPower}.
\end{equation}
From this and \eqref{eq:tipSpeedRatio}, the aerodynamic rotor torque ${M_r(t)\in\R}$ can be obtained ass
\begin{subequations}\label{eq:rotorTorque}
	\begin{align}
		M_r(t) =\frac{P_r(t)}{\omega_r(t)}&=m'\frac{V(t)^3}{\omega_r(t)}C_p(\lambda(t),u_\theta(t))\\
		&=\underbrace{m'R}_{m}V(t)^2\frac{C_p(\lambda(t),u_\theta(t))}{\lambda(t)}.
	\end{align}
\end{subequations}

\subsection{Drive train and gearbox subsystem}
The wind turbine is modeled with a rigid drive train and gearbox, i.e., the generator angular velocity is ${\omega_g(t) = N_g\omega_r(t)}$, where ${N_g\in\Rp}$ denotes the gearbox ratio.
The dynamics are given by
\begin{equation}
J\dot{\omega}_r(t) = M_r(t) - \underbrace{N_gu_M(t)}_{M_g(t)}\label{eq:torqueBalance},
\end{equation}
where ${u_M(t),M_g(t)\in\R}$ denote the generator torque on the generator and rotor side, respectively and ${J\in\Rp}$ the combined generator, rotor hub and blade inertia.

\subsection{Generator subsystem}
The generator converts mechanical into electrical power with efficiency $\eta\in(0,1)\subset\R$, i.e.,
\begin{equation}
P_e(t)=\eta\omega_g(t)u_M(t) = \eta N_g\omega_r(t)u_M(t).\label{eq:outputPower}
\end{equation}

\subsection{State model}
From \eqref{eq:tipSpeedRatio} and \eqref{eq:rotorTorque} to \eqref{eq:outputPower} a model with state $\omega_r(t)$, control input ${\mathbf{u}(t) = [u_\theta(t)\quad u_M(t)]^T}$ and uncertain input $V(t)$ of the form
\begin{subequations}\label{eq:contModel}
	\begin{equation}
		\textstyle\dot{\omega}_r(t) = \frac{m}{J}V(t)^2\frac{C_p\left(\llambda{t}, u_\theta(t)\right)}{\llambda{t}} - \frac{N_g}{J}u_M(t),\label{eq:dotOmega}
	\end{equation}
	can be obtained.
	The output is the electrical power, i.e.,
	\begin{equation}
		P_e(t) =\eta N_g\omega_r(t)u_M(t).
	\end{equation}
\end{subequations}
In what follows, a discrete-time model with sampling time ${T_s\in\Rp}$ is used, i.e., $\omega_r(t)$ turns into $\omega_r(kT_s)$, with ${k\in\mathbb{Z}}$.
For simplicity, $\omega_r(k)$ is shorthand for $\omega_r(kT_s)$.
Using the forward Euler method, the following discrete-time state model can be obtained from \eqref{eq:contModel}:
\begin{subequations}\label{eq:discModel}
	\begin{align}
		\omega_r(k&+1) = \omega_r(k)\nonumber\\
		&+T_s\textstyle{\left(\frac{m}{J}V(k)^2\frac{C_p\left(\llambda{k}, u_\theta(k)\right)}{\llambda{k}}-\frac{N_g}{J}u_M(k)\right)}\label{eq:discOmega}\\
		&P_e(k) = \eta N_g\omega_r(k)u_M(k).
	\end{align}
\end{subequations}
This model can now be used to derive a model predictive controller.


\section{MODEL PREDICTIVE CONTROLLER}
\label{sec:mpc}
In this section, an offset-free \ac{mpc} approach is presented.
It uses a model of a wind turbine based on \eqref{eq:discModel} to find an optimal control trajectory that minimizes a cost function subject to constraints.
First, a continuous piecewise-affine approximation of $C_p$, denoted as $\hat C_p$, is introduced.
Then, a wind forecast model, a bilinear reformulation for \eqref{eq:discOmega} and a disturbance observer are introduced and constraints presented.
Finally, an \ac{mpc} problem is formulated.

\subsection{Approximation of power coefficient}\label{subsec:pwa}
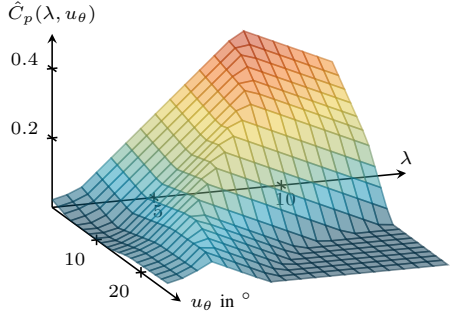
\begin{figure}[t]
	\centering

\tikzset{external/export next=false}
\begin{tikzpicture}[font=\scriptsize]
  \begin{axis}[%
    xlabel = {$\lambda$},
    ylabel = {$u_\theta$ in $^\circ$},
    zlabel = {$\hat C_p(\lambda, u_\theta)$},
    view = {20}{-30},
    axis lines = middle,
    xlabel style={anchor=south},
    ylabel style={anchor=west},
    zlabel style={anchor=south},
    xmax = 14.9,
    xmin = 1,
    ymax = 29,
    ymin = 0,
    zmax = 0.5,
    zmin = 0,
    restrict z to domain = 0:1,
    restrict y to domain = 0:26,
    restrict x to domain = 0:12,
    height = 5.6cm,
    width = 8cm,
    ]

    \addplot3[surf, mesh/cols=20, z buffer = auto, opacity=0.5] table [x=lambda, y=theta, z=Cp, col sep=comma] {figures/csvFiles/PWACp.csv} ;

  \end{axis}
\end{tikzpicture}
	\caption{Piecewise-affine approximation of $\hat C_p$ for $N_R=9$.}
	\label{fig:PWACp}
\end{figure}

The design of the \ac{mpc} scheme is based on \eqref{eq:discOmega}.
The latter contains the power coefficient $C_p(\lambda,u_\theta)$, which is a nonlinear function of $\lambda$ and $u_\theta$.
To obtain a computationally tractable \ac{mpc} formulation, a piecewise-affine approximation of the power coefficient is used.
For this, we define ${N_R\in\mathbb{N}}$ closed subsets $S_j$, ${j\in[1,N_R]\subset\mathbb{N}}$, in $\R^2$ of quadrilateral shape by systems of linear inequalities
\begin{equation}\label{eq:Regions}
    \mathbf{A}_j
        \begin{bmatrix}
            \lambda & u_\theta
        \end{bmatrix}^T
        \leq\mathbf{b}_j,
\end{equation}
where ${\mathbf{A}_j\in\R^{4\times2}}$ and ${\mathbf{b}_j\in\R^4}$.
We choose ${\mathbf{A}_j,\mathbf{b}_j}$ such that the corresponding sets $S_j$, ${j\in[1,N_R]}$ cover the relevant part of the $\lambda$-$u_\theta$-plane and any interior point of $S_j$, ${j\in[1,N_R]}$ does not belong to any subset $S_i$, ${i\neq j}$.
We then define a piecewise affine approximation of $C_p$:
\begin{equation}
    \hat C_p(\lambda,u_\theta)=a_j\lambda+b_ju_\theta+c_j\quad\text{if }[\lambda\quad u_\theta]^T\in S_j,
\end{equation}
where ${a_j,b_j,c_j}$ are chosen such that $\hat C_p$ is continuous over all subsets $S_j$.
An example of $\hat C_p$ is depicted in \cref{fig:PWACp}.

\begin{subequations}\label{eq:bigM}
Let the bounds ${\mathbf{M}\in\R^4}$ and ${m_C,M_C\in\R}$ be chosen such that ${\mathbf{A}_j[\lambda\quad u_\theta]^T\leq \mathbf{M}}$ and ${m_C\leq a_j\lambda+b_ju_\theta+c_j\leq M_C}$ for all ${j\in[1,N_R]}$.
Motivated by \cite{Bemporad_1999}, we use these bounds and additional binary variables $\delta_j\in\{0,1\}$ to obtain the following representation of the piecewise affine approximation
    \begin{align}
        \hat{C}_p &=\textstyle\sum_{j=1}^{N_R}\hat C_{p,j},\label{eq:sumCp}\\
        1&=\textstyle\sum_{j=1}^{N_R}\delta_j,\label{eq:sumRegions}\\
        \mathbf{A}_j
        \begin{bmatrix}
            \lambda &
            u_\theta
        \end{bmatrix}^T
            &\leq\delta_j\mathbf{b}_j+(1-\delta_j)\mathbf{M},\label{eq:minmaxLambdaTheta}\\
        \delta_jm_C&\leq\hat C_{p,j}\leq\delta_jM_C,\label{eq:minmaxCpj}\\
        m_C(1-\delta_j)\leq a_j\lambda&+b_ju_\theta+c_j-\hat C_{p,j}\leq M_C(1-\delta_j),\label{eq:minmaxAjBj}
    \end{align}
for all ${j\in[1,N_R]}$.
\end{subequations}
\begin{remark}
    One can distinguish between two cases for each subset ${j\in[1,N_R]}$:
    \begin{enumerate}
        \item   For ${\delta_{j}=0}$, \eqref{eq:minmaxCpj} becomes ${0\leq \hat C_{p,j}\leq0}$, i.e., ${\hat C_{p,j}=0}$.
                Equation \eqref{eq:minmaxAjBj} becomes ${m_c\leq a_j\lambda+b_ju_\theta+c_j\leq M_c}$ which does not impose any restrictions on $\lambda$ and $u_\theta$ by choice of $m_c$ and $M_c$.
                Moreover, the upper bound in \eqref{eq:minmaxLambdaTheta} becomes $\mathbf{M}$ which does not restrict ${[\lambda\quad u_\theta]^T}$ either.
        \item   For ${\delta_{j}=1}$, \eqref{eq:minmaxCpj} becomes ${m_C\leq\hat C_{p,j}\leq M_C}$ which does not impose any restrictions on ${\hat C_{p,j}}$.
                Moreover, \eqref{eq:minmaxAjBj} becomes equivalent to ${\hat C_{p,j}=a_{j}\lambda+b_{j}u_\theta+c_{j}}$.
                In \eqref{eq:minmaxLambdaTheta}, the upper bound becomes $\mathbf{b}_j$, which turns this inequality into \eqref{eq:Regions}.
                Equation \eqref{eq:sumRegions} leads to ${\delta_i=0\,\forall i\in[1,N_R]\backslash\{j\}}$, and \eqref{eq:sumCp} becomes ${\hat C_p=\hat C_{p,j}=a_j\lambda+b_ju_\theta+c_j}$.
    \end{enumerate}
\end{remark}

With $\hat C_p$, a control oriented version of \eqref{eq:discOmega} of the form
\begin{multline}\label{eq:discOmegaHatCp}
    \omega_r(n+1|k) = \omega_r(n|k)\\
    +T_s\textstyle{\left(\frac{m}{J}V(n|k)^2\frac{\hat C_p(n|k)}{\llambda{n|k}}-\frac{N_g}{J}u_M(n|k)\right)}
\end{multline}
can be derived $\forall n\in[k,k+N_p]$, where $N_p\in\mathbb{N}$ denotes the controller prediction horizon.
The notation $\omega_r(n|k)$ refers to a prediction for time $n$ performed at time $k$.
Equation \eqref{eq:discOmegaHatCp} can be used in the \ac{mpc} to predict the system behaviour into the future.

\subsection{Wind speed forecast}\label{subsec:windForecast}
To predict the system behaviour, a forecast of the wind speed $V$ is required.
In what follows, a widely employed persistence forecast \cite[sec. 6.4]{hans2021operation} is used, i.e.,
\begin{equation}\label{eq:windForecast}
    V(n|k) = V(k).
\end{equation}
$V(k)$ is assumed to be available through light detection and ranging (LIDAR) measurements or estimations (see, e.g., \cite{Hafidi_2012,Schreiber_2020, Nasrabad_2021}).
With \eqref{eq:windForecast}, \eqref{eq:discOmegaHatCp} becomes
\begin{multline}\label{eq:discOmegaHatCpWindForecast}
    \omega_r(n+1|k) = \omega_r(n|k)\\
    +T_s\textstyle{\left(\frac{m}{J}V(k)^2\frac{\hat C_p(n|k)}{R\frac{\omega_r(n|k)}{V(k)}}-\frac{N_g}{J}u_M(n|k)\right)}.
\end{multline}

\subsection{Bilinear reformulation}\label{subsec:bilinear}
The term
\begin{equation}
    \frac{\hat C_p(n|k)}{R\frac{\omega_r(n|k)}{V(k)}}\label{eq:zNonLinear}
\end{equation}
in \eqref{eq:discOmegaHatCpWindForecast} includes a division of decision variables.
Introducing $z(n|k)$ by the bilinear expression
\begin{equation}
    z(n|k)\omega_r(n|k)=\frac{V(k)}{R}\label{eq:auxiliaryEquality}
\end{equation}
allows to state \eqref{eq:zNonLinear} as $z(n|k)\hat C_p(n|k)$ which is bilinear as well.
With \eqref{eq:auxiliaryEquality}, the prediction of the state can be written as
\begin{multline}\label{eq:discOmegaHatCpWindForecastBilinear}
    \textstyle\omega_r(n+1|k) = \omega_r(n|k)\\
    +T_s\frac{mV(k)^2z(n|k)\hat C_p(n|k)-N_gu_M(n|k)}{J}.
\end{multline}
The advantage of \eqref{eq:auxiliaryEquality} and \eqref{eq:discOmegaHatCpWindForecastBilinear} over \eqref{eq:discOmegaHatCpWindForecast} is that they can be used in bilinear optimization problems which can be efficiently solved by a larger variety of available solvers.

\subsection{Offset-free control}\label{subsec:offsetFree}
Equation \eqref{eq:discOmegaHatCpWindForecastBilinear} is subject to model mismatch.
Therefore, power tracking would not be offset free.
Here, this mismatch is modelled by an input disturbance $d_\omega(n|k)$.
Estimating this disturbance eliminates the offset \cite{Pannocchia_2015}.
Here, a persistence forecast is used, i.e., 
\begin{equation}
    d_\omega(n|k) = d_\omega(k),
\end{equation}
where ${d_\omega(k\leq0)=0}$.
The signal is updated using the predicted $\omega_r(k+1|k)$ and the measured $\omega_r(k+1)$ via
\begin{equation}
    d_\omega(k+1) = d_\omega(k)+L\left(\omega_r(k+1)-\omega_r(k+1|k)\right),
\end{equation}
where $L\in\R$ is the Luenberger observer gain (see \cite{Pannocchia_2015}) which is found using a linearized wind turbine model \cite{Grapentin2022}.
Finally, the prediction used in the \ac{mpc} can be posed as
\begin{multline}\label{eq:discOmegaHatCpWindForecastBilinearOffsetFree}
    \textstyle\omega_r(n+1|k) = \omega_r(n|k)+d_\omega(k)\\
    +T_s\frac{mV(k)^2z(n|k)\hat C_p(n|k)-N_gu_M(n|k)}{J}.
\end{multline}

\subsection{Constraints}\label{subsec:Constraints}
\begin{subequations}\label{ineq:u}
The control input and its rate of change are bounded ${\forall k\in\mathbb{Z}}$, ${\forall n\in[k,k+N_p]}$, i.e.,
    \begin{align}
        \underline{\mathbf{u}}&\leq\mathbf{u}(n|k)\leq\overline{\mathbf{u}}\\
        \underline\Delta\mathbf{u}&\leq \mathbf{u}(n|k)-\mathbf{u}(n-1|k)\leq \overline\Delta\mathbf{u}
    \end{align}
\end{subequations}
with $\mathbf{u}(k-1|k)=\mathbf{u}(k-1)$ and bounds ${\underline{\mathbf{u}},\overline{\mathbf{u}}\in\R^2}$ and ${\underline\Delta\mathbf{u},\overline\Delta\mathbf{u}\in\R^2}$.
Additionally, to formulate a penalty for changes of $\mathbf{u}$, the constraint
\begin{equation}\label{ineq:uSoft}
    -\epsilon_{\Delta u}(n|k)\leq\mathbf{u}(n|k)-\mathbf{u}(n-1|k)\leq\epsilon_{\Delta u}(n|k),
\end{equation}
with $\epsilon_{\Delta u}(n|k)\in\R^2$ is introduced.
It can be used, for example, to reduce mechanical wear on the pitch actuators (see \cref{subsec:Problem}).
To ensure feasibility, soft constraints on the state of the form
\begin{equation}\label{ineq:omega}
    \underline\omega_r-\epsilon_\omega(n|k)\leq\omega_r(n|k)\leq\overline\omega_r+\epsilon_\omega(n|k),
\end{equation}
are employed where $\underline\omega_r,\overline\omega_r\in\R$ denote the lower and upper bound and $\epsilon_\omega(n|k)\in\R$ is a slack variable.
Both, $\epsilon_\omega$ and $\epsilon_{\Delta u}$ are nonnegative, i.e.,
\begin{subequations}\label{ineq:epsilon}
    \begin{align}
        0&\leq\epsilon_\omega(n|k)\label{ineq:epsilonOmega}\\
        0&\leq\epsilon_{\Delta u}(n|k)\label{ineq:epsilonDeltaU}.
    \end{align}
\end{subequations}
The output power has a lower bound of zero, as no power shall be drawn from the grid, and upper bound $P_e^d(k)$ which is an external input to the controller, i.e.,
\begin{equation}\label{ineq:power}
    0\leq \eta N_g\omega_r(n|k)u_M(n|k)\leq P_e^d(k).
\end{equation}
Thus, if enough wind power is available, then the output power should track $P_e^d(k)$.

\subsection{Problem formulation}\label{subsec:Problem}
All decision variables at prediction step $n$ are collected in
\begin{multline}
    \mathbf{x}(n|k) = [\omega_r(n|k)\quad\mathbf{u}(n|k)^T\\
    \hat C_p(n|k)\quad z(n|k)\quad\epsilon_\omega(n|k)\\
    \epsilon_{\Delta u}(n|k)^T\quad\delta_1(n|k)\ldots\delta_{N_R}(n|k)]^T.
\end{multline}
The decision variables over prediction horizon $N_p$ are collected in
\begin{equation}
    \resizebox{0.89\hsize}{!}{$\mathbf{X}(k) = [\mathbf{x}(k|k)^T\ldots\,\mathbf{x}(k+N_p-1|k)^T\vphantom{x(k)^T}\omega_r(k+N_p|k)]^T.$}
\end{equation}
The cost function to be minimized at step $k$ is
\begin{multline}\label{eq:cost}
        \ell\left(\mathbf{X}(k)\right)=\textstyle\sum_{n=k}^{k+N_p}\left(\left(-\eta N_g\omega_r(n|k)u_M(n|k)\right.\right.\\
        \left.\left.+q_\omega\epsilon_\omega(n|k)+\mathbf{q}_{\Delta u}^T\epsilon_{\Delta u}(n|k)\right)\gamma^{n-k}\right).
\end{multline}
Here, $q_\omega\in\Rp$ and $\mathbf{q}_{\Delta u}\in\R^2$, $\mathbf{q}_{\Delta u}>0$ are weights.
Moreover, $\gamma\in(0,1)$ is used to emphasize decisions in the near future.
Hence control aims at rewarding output power $P_e(k)=\eta N_g\omega_r(k)u_M(k)$ and at penalizing $\epsilon_\omega(k)$ and $\epsilon_{\Delta u}$.
Together with \eqref{ineq:power}, \eqref{eq:cost} yields a controller which maximizes $P_e(k)$ while ensuring that $P_e^d(k)$ is not exceeded.
Finally, the following \ac{mpc} problem for a wind turbine can be formulated.

\begin{problem}\label{eq:optimizationProgram}
    \begin{subequations}
        \begin{equation*}
            \min_{\mathbf{X}(k)}\quad\ell\left(\mathbf{X}(k)\right)
        \end{equation*}
        subject to
        \begin{multline*}
            \textstyle\omega_r(n+1|k) = \omega_r(n|k)+d_\omega(k) \\
            +T_s\frac{mV(k)^2z(n|k)\hat C_p(n|k)-N_gu_M(n|k)}{J},
        \end{multline*}
    \end{subequations}
    as well as \cref{eq:bigM,eq:auxiliaryEquality,ineq:u,ineq:uSoft,ineq:omega,ineq:epsilon,ineq:power} ${\forall n\in[k,k+N_p]}$, with measured ${\omega_r(k|k) = \omega_r(k)}$ and ${u(k-1|k)=u(k-1)}$.
\end{problem}

The solution for $\mathbf{u}(k|k)$ obtained by solving \cref{eq:optimizationProgram} is used to actuate the system at time $k$.
At time $k+1$, a new measurement $\omega_r(k+1|k+1)=\omega_r(k+1)$ is obtained, $d_\omega(k+1)$ and $V(k+1)$ are updated and the optimization process is repeated.


\section{CASE STUDY}
\label{sec:caseStudy}

In this section, our \ac{mpc} scheme is compared to the baseline controller from \cite{Grapentin2022}.
For this comparison the IEA 3.4-MW land-based wind turbine from \cite{RWT} is used.
The parameters for the \ac{mpc} can be found in \cref{tab:mpcPara}.
\begin{table}[b]
    \centering
    \caption{\ac{mpc} PARAMETERS}
    \label{tab:mpcPara}
    \begin{tabular}{cl}
        \toprule
        Parameter & Value \\
        \midrule
        $\underline{\omega}_r$ & $\SI{0.3979}{\radian\per\second}$ \\
        $\overline{\omega}_r$ & $\SI{1.2305}{\radian\per\second}$ \\
        $\mathbf{\underline{u}}$ & $[\SI{0.5263}{\degree} \quad \SI{0}{\kilo\newton\meter}]^T$ \\
        $\mathbf{\overline{u}}$ & $[\SI{26}{\degree} \quad \SI{31.5751}{\kilo\newton\meter}]^T$ \\
        $\Delta\mathbf{\underline{u}}$ & $[-\SI{21}{\degree} \quad -\SI{45}{\kilo\newton\meter}]^T$ \\
        $\Delta\mathbf{\overline{u}}$ & $[\SI{21}{\degree} \quad \SI{45}{\kilo\newton\meter}]^T$ \\
        $L$ & $0.5$ \\
        $T_s$ & $\SI{3}{\second}$ \\
        $N_p$ & $6$ \\
        \midrule
        $q_\omega$ & $10^5\,\si{\second\per\radian}$ \\
        $q_{\Delta u}$ & $[\si{1\per\degree}\quad\si{1\per\kilo\newton\meter}]^T$ \\
        $\gamma$ & $0.99$ \\
        \bottomrule
    \end{tabular}
\end{table}
Additionally, the parameters for $\hat C_p$ can be found in \cite{zenodoMPC}.
The \ac{mpc} algorithm is implemented in MATLAB \cite{MATLAB:2021} using the Yalmip toolbox \cite{Lofberg2004}.
Gurobi \cite{gurobi} is used to numerically solve \cref{eq:optimizationProgram}.
Unit tests have been deployed to ensure that the obtained results are correct.
All simulations are conducted on a small server with an Intel\textsuperscript{\textregistered} Xeon\textsuperscript{\textregistered} E5-1620 v2 processor @$\SI{3.70}{\giga\hertz}$ with 4 CPU cores and $\SI{32}{\giga\byte}$ RAM.
A sampling time $T_s=\SI{3}{\second}$ and a prediction horizon of $6$ samples are considered.
For each simulation, the first $\SI{2}{\minute}$ are omitted to disregard startup effects of the simulation tool OpenFAST.
All simulations were conducted with nonzero turbulence intensities (see \cite{MACEACHERN2018665}).
In detail, a turbulence intensity of $\SI{9}{\percent}$ was considered.

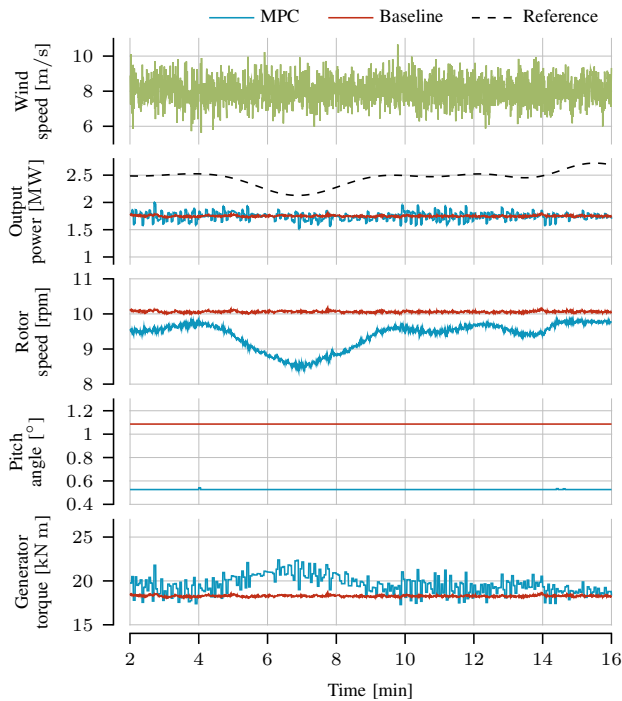
\begin{figure}[t]
    \centering

\tikzset{external/export next=false}

\begin{tikzpicture}[font=\scriptsize]
  \begin{axis}[%
    myPlot,
    height = 3cm,
    ylabel = {\begin{tabular}{c}Wind\\[-.1em]speed [\si{\meter\per\second}]\end{tabular}},
    grid = both,
    major grid style = {line width=.2pt, draw=gray!50},
    ymin = 5,
    ymax = 11,
    xmin = 2,
    xmax = 16,
    xmajorticks = false,
    x axis line style = {white},
    clip = true,
    legend columns=3,
    legend style={
      at={(1,1)},
      yshift = .2,
      anchor=south east,
      draw=none,
      fill=none,
      legend cell align=left,
      /tikz/every even column/.append style={column sep=0.3cm},
      legend entries = {\ac{mpc}, Baseline, Reference},
      },
    ]

    \addplot[color=vgLightBlue] coordinates {(-1,-1)};
    \addplot[color=vgRed] coordinates {(-1,-1)};
    \addplot[color=black, dashed] coordinates {(-1,-1)};
    \addplot[vgGreen, line width=0.6pt] table [x expr=\thisrow{time}/60, y ={windSpeed}, col sep=comma] {figures/csvFiles/powerMax.csv};
  \end{axis}

  \begin{axis}[%
    myPlot,
    height = 3cm,
    ylabel = {\begin{tabular}{c}Output\\[-.1em]power [\si{\mega\watt}]\end{tabular}},
    grid = both,
    major grid style = {line width=.2pt, draw=gray!50},
    yshift = -1.6cm,
    ymin = 0.87,
    ymax = 2.8,
    xmin = 2,
    xmax = 16,
    xmajorticks = false,
    x axis line style = {white},
    clip = true,
    ]
    
    \addplot[vgLightBlue, line width=0.6pt] table [x expr=\thisrow{time}/60, y ={outputMPC}, col sep=comma] {figures/csvFiles/powerMax.csv};
    \addplot[vgRed, line width=0.6pt] table [x expr=\thisrow{time}/60, y ={outputBLC}, col sep=comma] {figures/csvFiles/powerMax.csv};
    \addplot[black, dashed, line width=0.6pt] table [x expr=\thisrow{time}/60, y ={outputRef}, col sep=comma] {figures/csvFiles/powerMax.csv};
    
  \end{axis}

  \begin{axis}[%
    myPlot,
    height = 3cm,
    ylabel = {\begin{tabular}{c}Rotor\\[-.1em]speed [rpm]\end{tabular}},
    grid = both,
    major grid style = {line width=.2pt, draw=gray!50},
    yshift = -3.2cm,
    ymin = 8,
    ymax = 11,
    xmin = 2,
    xmax = 16,
    xmajorticks = false,
    x axis line style = {white},
    clip = true,
    ]
    
    \addplot[vgLightBlue, line width=0.6pt] table [x expr=\thisrow{time}/60, y expr=\thisrow{rotorSpeedMPC}, col sep=comma] {figures/csvFiles/powerMax.csv};
    \addplot[vgRed, line width=0.6pt] table [x expr=\thisrow{time}/60, y expr=\thisrow{rotorSpeedBLC}, col sep=comma] {figures/csvFiles/powerMax.csv};
  
  \end{axis}

  \begin{axis}[%
    myPlot,
    height = 3cm,
    ylabel = {\begin{tabular}{c}Pitch\\[-.1em]angle [\si{\degree}]\end{tabular}},
    grid = both,
    major grid style = {line width=.2pt, draw=gray!50},
    yshift = -4.8cm,
    ymin = 0.4,
    ymax = 1.3,
    xmin = 2,
    xmax = 16,
    xmajorticks = false,
    x axis line style = {white},
    ]
    
    \addplot[vgLightBlue, line width=0.6pt] table [x expr=\thisrow{time}/60, y expr=\thisrow{pitchAngleMPC}, col sep=comma] {figures/csvFiles/powerMax.csv};
    \addplot[vgRed, line width=0.6pt] table [x expr=\thisrow{time}/60, y expr=\thisrow{pitchAngleBLC}, col sep=comma] {figures/csvFiles/powerMax.csv};
  \end{axis}

  \begin{axis}[%
    myPlot,
    height = 3cm,
    title style = {yshift = -.7cm, fill=white},
    xlabel = {Time [min]},
    ylabel = {\begin{tabular}{c}Generator\\[-.1em]torque [\si{\kilo\newton\meter}]\end{tabular}},
    grid = both,
    major grid style = {line width=.2pt, draw=gray!50},
    yshift = -6.4cm,
    ymin = 15,
    ymax = 27,
    xmin = 2,
    xmax = 16,
    ]

    \addplot[vgLightBlue, line width=0.6pt] table [x expr=\thisrow{time}/60, y expr=\thisrow{generatorTorqueMPC}, col sep=comma] {figures/csvFiles/powerMax.csv};
    \addplot[vgRed, line width=0.6pt] table [x expr=\thisrow{time}/60, y expr=\thisrow{generatorTorqueBLC}, col sep=comma] {figures/csvFiles/powerMax.csv};

  \end{axis}
\end{tikzpicture}
    \caption{Power maximization for a \SI{8}{\meter\per\second}.}
    \label{fig:powerMax}
\end{figure}

In \cref{fig:powerMax}, simulation results obtained with the \ac{mpc} and the baseline controller are depicted for an average wind speed of $\SI{8}{\meter\per\second}$.
For this wind speed, the available power is below the demanded one and thus, the generated power is maximized.
Note, that while the baseline controller keeps the rotor speed at a constant level, the \ac{mpc} changes the rotor speed with the power reference.
However, this has no effect on the power maximization as the generator torque counteracts this effect.
Moreover, the pitch angle is kept constant for both controllers to keep the power coefficient at its maximum value.
From \cref{fig:powerCoefficientData}, it becomes apparent that the power coefficient has a maximum plateau rather than a singular maximum point.
Hence, different pitch angles may may lead to the same maximum power.
The general behaviour of the \ac{mpc} in this operating region is currently being further investigated.

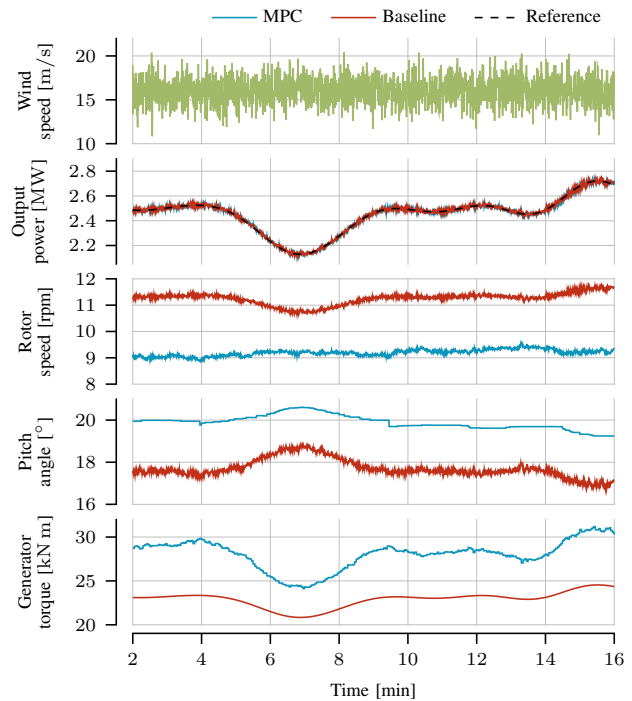
\begin{figure}
    \centering

\tikzset{external/export next=false}

\begin{tikzpicture}[font=\scriptsize]
  \begin{axis}[%
    myPlot,
    height = 3cm,
    ylabel = {\begin{tabular}{c}Wind\\[-.1em]speed [\si{\meter\per\second}]\end{tabular}},
    grid = both,
    major grid style = {line width=.2pt, draw=gray!50},
    ymin = 10,
    ymax = 22,
    xmin = 2,
    xmax = 16,
    xmajorticks = false,
    x axis line style = {white},
    clip = true,
    legend columns=3,
    legend style={
      at={(1,1)},
      yshift = .2,
      anchor=south east,
      draw=none,
      fill=none,
      legend cell align=left,
      /tikz/every even column/.append style={column sep=0.3cm},
      legend entries = {\ac{mpc}, Baseline, Reference},
      },
    ]

    \addplot[color=vgLightBlue] coordinates {(-1,-1)};
    \addplot[color=vgRed] coordinates {(-1,-1)};
    \addplot[color=black, dashed] coordinates {(-1,-1)};
    \addplot[vgGreen, line width=0.6pt] table [x expr=\thisrow{time}/60, y ={windSpeed}, col sep=comma] {figures/csvFiles/powerTracking.csv};
  \end{axis}

  \begin{axis}[%
    myPlot,
    height = 3cm,
    ylabel = {\begin{tabular}{c}Output\\[-.1em]power [\si{\mega\watt}]\end{tabular}},
    grid = both,
    major grid style = {line width=.2pt, draw=gray!50},
    yshift = -1.6cm,
    ymin = 2.05,
    ymax = 2.9,
    xmin = 2,
    xmax = 16,
    xmajorticks = false,
    x axis line style = {white},
    ]
    
    \addplot[vgLightBlue, line width=0.6pt] table [x expr=\thisrow{time}/60, y ={outputMPC}, col sep=comma] {figures/csvFiles/powerTracking.csv};
    \addplot[vgRed, line width=0.6pt] table [x expr=\thisrow{time}/60, y ={outputBLC}, col sep=comma] {figures/csvFiles/powerTracking.csv};
    \addplot[color=black, dashed, line width = 0.6] table [x expr=\thisrow{time}/60, y ={reference}, col sep=comma] {figures/csvFiles/powerTracking.csv};
    
  \end{axis}

  \begin{axis}[%
    myPlot,
    height = 3cm,
    ylabel = {\begin{tabular}{c}Rotor\\[-.1em]speed [rpm]\end{tabular}},
    grid = both,
    major grid style = {line width=.2pt, draw=gray!50},
    yshift = -3.2cm,
    ymin = 8,
    ymax = 12,
    xmin = 2,
    xmax = 16,
    xmajorticks = false,
    x axis line style = {white},
    legend columns = -1,
    clip = true,
    ]
    
    \addplot[vgLightBlue, line width=0.6pt] table [x expr=\thisrow{time}/60, y expr=\thisrow{rotorSpeedMPC}, col sep=comma] {figures/csvFiles/powerTracking.csv};
    \addplot[vgRed, line width=0.6pt] table [x expr=\thisrow{time}/60, y expr=\thisrow{rotorSpeedBLC}, col sep=comma] {figures/csvFiles/powerTracking.csv};
  
  \end{axis}

  \begin{axis}[%
    myPlot,
    height = 3cm,
    ylabel = {\begin{tabular}{c}Pitch\\[-.1em]angle [\si{\degree}]\end{tabular}},
    grid = both,
    major grid style = {line width=.2pt, draw=gray!50},
    yshift = -4.8cm,
    ymin = 16,
    ymax = 21,
    xmin = 2,
    xmax = 16,
    xmajorticks = false,
    x axis line style = {white},
    legend pos = south west,
    legend columns = -1,
    ]
    
    \addplot[vgLightBlue, line width=0.6pt] table [x expr=\thisrow{time}/60, y expr=\thisrow{pitchAngleMPC}, col sep=comma] {figures/csvFiles/powerTracking.csv};
    \addplot[vgRed, line width=0.6pt] table [x expr=\thisrow{time}/60, y expr=\thisrow{pitchAngleBLC}, col sep=comma] {figures/csvFiles/powerTracking.csv};
  \end{axis}

  \begin{axis}[%
    myPlot,
    height = 3cm,
    title style = {yshift = -.7cm, fill=white},
    xlabel = {Time [min]},
    ylabel = {\begin{tabular}{c}Generator\\[-.1em]torque [\si{\kilo\newton\meter}]\end{tabular}},
    grid = both,
    major grid style = {line width=.2pt, draw=gray!50},
    yshift = -6.4cm,
    ymin = 20,
    ymax = 32,
    xmin = 2,
    xmax = 16,
    ]

    \addplot[vgLightBlue, line width=0.6pt] table [x expr=\thisrow{time}/60, y expr=\thisrow{generatorTorqueMPC}, col sep=comma] {figures/csvFiles/powerTracking.csv};
    \addplot[vgRed, line width=0.6pt] table [x expr=\thisrow{time}/60, y expr=\thisrow{generatorTorqueBLC}, col sep=comma] {figures/csvFiles/powerTracking.csv};

  \end{axis}
\end{tikzpicture}
    \caption{Power tracking for an average wind speed of \SI{16}{\meter\per\second}.}
    \label{fig:powerTracking}
\end{figure}

In \cref{fig:powerTracking}, simulation results for an average wind speed of $\SI{16}{\meter\per\second}$ are shown.
For this wind speed, the available power exceeds the demanded one and thus needs to be curtailed.
Both controllers track the reference well.
One notable difference is that the \ac{mpc} approach employs less control action which becomes evident when observing the pitch angle.
Consequently, the wear on the pitch actuators is reduced which may reduce long term maintenance cost.
Moreover, the \ac{mpc} approach employs a higher generator torque yielding a lower rotor speed, which reduces noise.

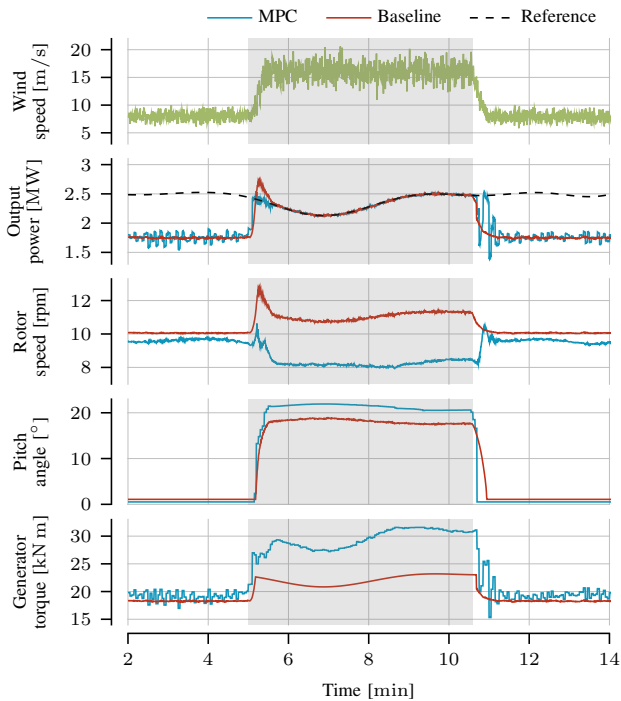
\begin{figure}[t]
    \centering

\tikzset{external/export next=false}

\newcommand{\xmin}{2}
\newcommand{\xmax}{14}
\newcommand{\yminW}{3}
\newcommand{\ymaxW}{22}
\newcommand{\xup}{5}
\newcommand{\xdown}{10.6}

\begin{tikzpicture}[font=\scriptsize]
  \begin{axis}[%
    myPlot,
    ylabel = {\begin{tabular}{c}Wind\\[-0.1em]speed [\si{\meter\per\second}]\end{tabular}},
    height = 3cm,
    grid = both,
    major grid style = {line width=.2pt, draw=gray!50},
    ymin = \yminW,
    ymax = \ymaxW,
    xmin = \xmin,
    xmax = \xmax,
    xmajorticks = false,
    x axis line style = {white},
    clip = true,
    legend columns=3,
    legend style={
      at={(1,1)},
      yshift = .2,
      anchor=south east,
      draw=none,
      fill=none,
      legend cell align=left,
      /tikz/every even column/.append style={column sep=0.3cm},
      legend entries = {\ac{mpc}, Baseline, Reference},
      },
    ]
    
    \addplot[color=vgLightBlue] coordinates {(-1,-1)};
    \addplot[color=vgRed] coordinates {(-1,-1)};
    \addplot[color=black, dashed] coordinates {(-1,-1)};

    \addplot[color=vgGreen] table [x=time, y=wind, col sep=comma] {figures/csvFiles/powerTrackingSwitch.csv};


    \fill [color=gray, draw=none, opacity=0.2] (\xup, \yminW) rectangle (\xdown, \ymaxW) node {};
  \end{axis}
  
  \newcommand{\yminP}{1.3}
  \newcommand{\ymaxP}{3.1}
  \begin{axis}[%
    myPlot,
    ylabel = {\begin{tabular}{c}Output\\[-.1em]power [\si{\mega\watt}]\end{tabular}},
    height = 3cm,
    grid = both,
    major grid style = {line width=.2pt, draw=gray!50},
    yshift = -1.6cm,
    ymin = \yminP,
    ymax = \ymaxP,
    xmin = \xmin,
    xmax = \xmax,
    xmajorticks = false,
    x axis line style = {white},
    ]
    \addplot[color=vgLightBlue] table [x=time, y=powerMPC, col sep=comma] {figures/csvFiles/powerTrackingSwitch.csv};
    \addplot[color=vgRed] table [x=time, y=powerBLC, col sep=comma] {figures/csvFiles/powerTrackingSwitch.csv};
    \addplot[color=black, dashed] table [x=time, y=powerRef, col sep=comma] {figures/csvFiles/powerTrackingSwitch.csv};
    
    \fill [color=gray, draw=none, opacity=0.2] (\xup, \yminP) rectangle (\xdown, \ymaxP) node {};
  \end{axis}

  \newcommand{\yminS}{7}
  \newcommand{\ymaxS}{13.3}
  \begin{axis}[%
    myPlot,
    ylabel = {\begin{tabular}{c}Rotor\\[-0.1em]speed [rpm]\end{tabular}},
    grid = both,
    height = 3cm,
    major grid style = {line width=.2pt, draw=gray!50},
    yshift = -3.2cm,
    ymin = \yminS,
    ymax = \ymaxS,
    xmin = \xmin,
    xmax = \xmax,
    xmajorticks = false,
    x axis line style = {white},
    ]
    \addplot[color=vgLightBlue] table [x=time, y=rotorSpeedMPC, col sep=comma] {figures/csvFiles/powerTrackingSwitch.csv};
    \addplot[color=vgRed] table [x=time, y=rotorSpeedBLC, col sep=comma] {figures/csvFiles/powerTrackingSwitch.csv};
    
    \fill [color=gray, draw=none, opacity=0.2] (\xup, \yminS) rectangle (\xdown, \ymaxS) node {};
  \end{axis}

  \newcommand{\yminA}{0}
  \newcommand{\ymaxA}{23}
  \begin{axis}[%
    myPlot,
    ylabel = {\begin{tabular}{c}Pitch\\[-.1em]angle [\si{\degree}]\end{tabular}},
    height = 3cm,
    grid = both,
    major grid style = {line width=.2pt, draw=gray!50},
    yshift = -4.8cm,
    ymin = \yminA,
    ymax = \ymaxA,
    xmin = \xmin,
    xmax = \xmax,
    xmajorticks = false,
    x axis line style = {white},
    ]
    \addplot[color=vgLightBlue] table [x=time, y=pitchAngleMPC, col sep=comma] {figures/csvFiles/powerTrackingSwitch.csv};
    \addplot[color=vgRed] table [x=time, y=pitchAngleBLC, col sep=comma] {figures/csvFiles/powerTrackingSwitch.csv};
    
    \fill [color=gray, draw=none, opacity=0.2] (\xup, \yminA) rectangle (\xdown, \ymaxA) node {};
  \end{axis} 

  \newcommand{\yminT}{14}
  \newcommand{\ymaxT}{33}
  \begin{axis}[%
    myPlot,
    xlabel = {Time [\si{\minute}]},
    ylabel = {\begin{tabular}{c}Generator\\[-.1em]torque [\si{\kilo\newton\meter}]\end{tabular}},
    height = 3cm,
    grid = both,
    major grid style = {line width=.2pt, draw=gray!50},
    yshift = -6.4cm,
    ymin = \yminT,
    ymax = \ymaxT,
    xmin = \xmin,
    xmax = \xmax,
    ]
    \addplot[color=vgLightBlue] table [x=time, y=generatorTorqueMPC, col sep=comma] {figures/csvFiles/powerTrackingSwitch.csv};
    \addplot[color=vgRed] table [x=time, y=generatorTorqueBLC, col sep=comma] {figures/csvFiles/powerTrackingSwitch.csv};
      
    \fill [color=gray, draw=none, opacity=0.2] (\xup, \yminT) rectangle (\xdown, \ymaxT) node {};
  \end{axis}

\end{tikzpicture}
    \caption{Controller performance for varying wind speeds.}
    \label{fig:powerTrackingSwitch}
\end{figure}

In \cref{fig:powerTrackingSwitch}, an operation where the average wind speed transitions back and forth between $6$ and $\SI{16}{\meter\per\second}$ is shown.
Both controllers are maximizing the output power for lower wind speeds which do not allow a generation of the reference power.
At higher wind speeds, both controllers track the reference power. 
However at changing wind conditions, the baseline controller yields an undesired overshoot.
The \ac{mpc} on the contrary, leads to a better reference tracking at this point.
When the wind speed decreases both controllers try to maximize the generated power.
Briefly, the \ac{mpc} approach tries to keep generating the reference power before reaching a new operating point.
This different approach leads to higher energy production as in the depicted period from $\SI{10.5}{\minute}$ to $\SI{12}{\minute}$, the turbine generates roughly $\SI{5.6}{\percent}$ more energy with the \ac{mpc} approach.

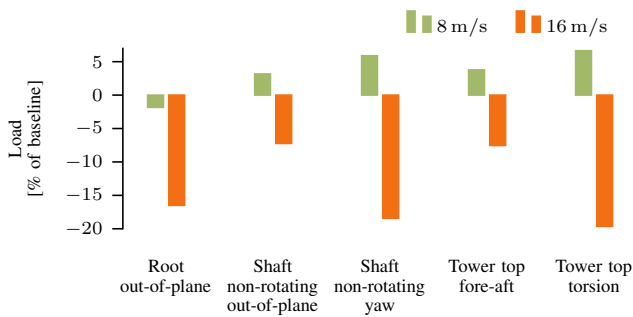
\begin{figure}
    \centering

\tikzset{external/export next=false}

\begin{tikzpicture}[font=\scriptsize]
  \begin{axis}[%
    myPlot,
    ybar,
    bar width = 6pt,
    ylabel={\begin{tabular}{c}Load\\[-.1em][\% of baseline]\end{tabular}},
    ytick = {-20,-15,-10,-5,0,5},
    xtick = {1, 2, 3, 4, 5},
    xticklabels = {
      \begin{tabular}{c}Root\\[-.1em]out-of-plane\end{tabular},
      \begin{tabular}{c}Shaft\\[-.1em]non-rotating\\[-.1em]out-of-plane\end{tabular},
      \begin{tabular}{c}Shaft\\[-.1em]non-rotating\\[-.1em]yaw\end{tabular},
      \begin{tabular}{c}Tower top\\[-.1em]fore-aft\end{tabular},
      \begin{tabular}{c}Tower top\\[-.1em]torsion\end{tabular}
    },
    ymin = -20,
    ymax = 7,
    xmin = 0.75,
    xmax = 5.25,
    x axis line style = {draw = none},
    x tick style = {draw = none},
    clip = true,
    legend columns=2,
    legend style={
      at={(1,1)},
      yshift = .2,
      anchor=south east,
      draw=none,
      fill=none,
      legend cell align=left,
      /tikz/every even column/.append style={column sep=0.3cm}
      },
    every axis plot/.append style = {
      fill
    }
    ]

    \addplot[vgGreen]   table[x = {x}, y = {y08}, col sep=comma] {figures/csvFiles/damageEquivalentLoads.csv};
    \addplot[vgOrange]  table[x = {x}, y = {y16}, col sep=comma] {figures/csvFiles/damageEquivalentLoads.csv};

    \legend{$\SI{8}{\meter\per\second}$,$\SI{16}{\meter\per\second}$};

  \end{axis}

\end{tikzpicture}
    \caption{\Aclp{del} at different wind speeds.}
    \label{fig:del}
\end{figure}

\cref{fig:del} shows the relative change for different \acp{del} when using the \ac{mpc} approach compared to the baseline controller.
They are computed using the rainflow algorithm \cite{dowling1971fatigue} and estimated for a 20 years life period \cite{burton2011wind}.
While some \acp{del} slightly increase at lower wind speeds, all of them significantly decrease at higher wind speeds when the \ac{mpc} approach is employed.
Therefore, the \ac{mpc} approach can reduce mechanical wear and thus replacement costs for operations with higher wind speeds.

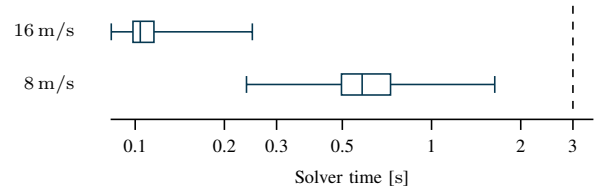
\begin{figure}
    \centering

\tikzset{external/export next=false}
  \begin{tikzpicture}[font=\scriptsize]
    \begin{axis}[
      myPlot,
      height = 3cm,
      boxplot,
      xmode = log,
      xlabel = {Solver time [\si{\second}]},
      xtick = {0.1,0.2,0.3,0.5,1,2,3,5},
      xticklabels = {0.1,0.2,0.3,0.5,1,2,3,5},
      ytick = {1,2},
      yticklabels = {$\SI{8}{\meter\per\second}$,$\SI{16}{\meter\per\second}$},
      y axis line style = {draw = none},
      y tick style = {draw = none},
      clip = true,
      xmax = 3.5,
      ymin = 0.5,
      ymax = 2.5,
      ]
      
    \addplot[vgDarkBlue] table [y = solverTime08, col sep=comma] {figures/csvFiles/solverTime.csv};
    \addplot[vgDarkBlue] table [y = solverTime16, col sep=comma] {figures/csvFiles/solverTime.csv};

    \draw[dashed] (axis cs:3,-10) -- (axis cs:3,10);
  \end{axis}
\end{tikzpicture}
    \caption{
        Distribution of solve times at different average wind speeds.
        The dashed line marks the sampling time of the \ac{mpc} approach.
    }
    \label{fig:solverTime}
\end{figure}

Finally, the solver time of the \ac{mpc} approach is evaluated.
\cref{fig:solverTime} depicts boxplots of the solver times for simulations with different wind speeds.
It can be observed that the solver time is smaller for higher wind speeds.
However at all times, it stays below the sampling time of $\SI{3}{\second}$ which renders the controller applicable for wind turbine applications.


\section{CONCLUSION}\label{sec:conclusion}

In this paper, a model predictive wind turbine controller with a piecewise-affine power coefficient approximation was presented.
It was shown that this approach favourably compares with state-of-the-art controllers in terms of power tracking accuracy, energy production and load reduction.

The developed MPC scheme provides a good basis for a variety of possible future extensions:
the controller can, for example, be extended to account explicitly for \aclp{del} by incorporating them in its cost function.
A persistent feasibility analysis shall be performed to ensure safe real-world application of the controller.
Additionally, the complexity of the \ac{mpc} scheme shall be reduced to decrease the solver time, which enables a larger prediction horizon.
Making the \ac{mpc} compute references and moving the actual control to a low layer controller could achieve this.
Another approach could be a modified construction of $\hat C_p$ such that less probable areas in the $\lambda-u_\theta$-plane are approximated with less subsets and more probable areas with more subsets.

\bibliographystyle{IEEEtran}
\bibliography{literature}

\end{document}